\documentclass[preprint, superscriptaddress]{revtex4}

\usepackage{graphicx}
\usepackage[ansinew]{inputenc}
\usepackage{array}
\usepackage{color}
\usepackage{amsmath}
\usepackage{amsxtra}
\usepackage{amstext}
\usepackage{amssymb}

\usepackage{latexsym}
\usepackage{dsfont}

\begin{document}
\renewcommand{\figurename}{Figure}

\title{STM Spectroscopy of ultra-flat graphene on hexagonal boron nitride}

\author{Jiamin Xue}
\affiliation{Department of Physics, University of Arizona, Tucson, AZ 85721 USA.}
\author{Javier Sanchez-Yamagishi}
\author{Danny Bulmash}
\affiliation{Department of Physics, Massachusetts Institute of Technology, Cambridge, MA 02138 USA.}
\author{Philippe Jacquod}
\affiliation{Department of Physics, University of Arizona, Tucson, AZ 85721 USA.}
\affiliation{College of Optical Sciences, University of Arizona, Tucson, AZ 85721 USA.}
\author{Aparna Deshpande}
\affiliation{Department of Physics, University of Arizona, Tucson, AZ 85721 USA.}
\author{K. Watanabe}
\author{T. Taniguchi}
\affiliation{Advanced Materials Laboratory, National Institute for Materials Science, 1-1 Namiki, Tsukuba 305-0044, Japan}
\author{Pablo Jarillo-Herrero}
\affiliation{Department of Physics, Massachusetts Institute of Technology, Cambridge, MA 02138 USA.}
\author{Brian J. LeRoy}
\email{leroy@physics.arizona.edu}
\affiliation{Department of Physics, University of Arizona, Tucson, AZ 85721 USA.}

\date{\today}

\begin{abstract}
Graphene has demonstrated great promise for future electronics technology as well as fundamental physics applications because of its linear energy-momentum dispersion relations which cross at the Dirac point\cite{Geim2007,CastroNeto2009}.  However, accessing the physics of the low density region at the Dirac point has been difficult because of the presence of disorder which leaves the graphene with local microscopic electron and hole puddles\cite{Martin2008, Deshpande2009, Zhang2009}, resulting in a finite density of carriers even at the charge neutrality point.  Efforts have been made to reduce the disorder by suspending graphene, leading to fabrication challenges and delicate devices which make local spectroscopic measurements difficult\cite{Du2008,Bolotin2008}.  Recently, it has been shown that placing graphene on hexagonal boron nitride (hBN) yields improved device performance\cite{Dean2010}.  In this letter, we use scanning tunneling microscopy to show that graphene conforms to hBN, as evidenced by the presence of Moir\'{e} patterns in the topographic images.  However, contrary to recent predictions\cite{Giovannetti2007, Slawinska2010}, this conformation does not lead to a sizable band gap due to the misalignment of the lattices.  Moreover, local spectroscopy measurements demonstrate that the electron-hole charge fluctuations are reduced by two orders of magnitude as compared to those on silicon oxide.  This leads to charge fluctuations which are as small as in suspended graphene\cite{Du2008}, opening up Dirac point physics to more diverse experiments than are possible on freestanding devices. 
\end{abstract}\maketitle

Graphene was first isolated on silicon dioxide because of the ability to image monolayer regions using an optical microscope\cite{Novoselov2005}.  However, the electronic properties of SiO$_2$ are not ideal for graphene because of its high roughness and trapped charges in the oxide.  These impurity-induced charge traps tend to cause the graphene to electronically break up into electron and hole doped regions at low charge density which both limit device performance and make the Dirac point physics inaccessible\cite{Martin2008, Deshpande2009, Zhang2009, Chen2008, Rossi2008}.  In order to create devices with less puddles, the substrate must be removed or changed.  One possibility to get rid of substrate interactions is to suspend graphene\cite{Du2008,Bolotin2008}, as shown by the drastic improvement in mobility which has enabled the observation of the fractional quantum Hall effect in suspended devices\cite{Du2009, Bolotin2009}.  However, the freely suspended monolayers are very delicate, leading to fabrication difficulties as well as strain\cite{Bao2009}.  Because of these difficulties, there have been no STM spectroscopy measurements of the local electronic properties of suspended graphene devices.  All of this points to the need for new substrates that offer mechanical support to the graphene without interfering with its electrical properties.  Recently, such a candidate substrate has been found with the demonstration of high-quality graphene devices on hexagonal boron nitride (hBN)\cite{Dean2010}.  Hexagonal boron nitride has the same atomic structure as graphene, but with a 1.8\% longer lattice constant\cite{Liu2003}, and shares many similar properties with graphene except that it is a wide-bandgap electric insulator\cite{Watanabe2004}.  The planar structure of hBN cleaves into an ultra-flat surface and the ionic bonding of hBN should leave it free of dangling bonds and charge traps at the surface resulting in less induced electron-hole puddles in graphene.  Indeed, graphene on hBN devices exhibit the highest mobility reported on any substrate, as well as narrow Dirac peak resistance widths, indicating reduced disorder and charge inhomogeneity\cite{Dean2010}.

\begin{figure}[h]
\includegraphics[]{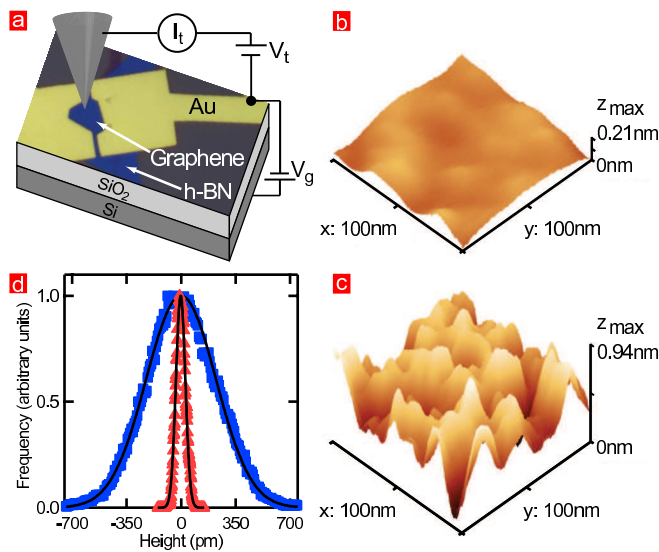}
\caption{Schematic device setup and topography comparison of graphene on hBN and SiO$_2$. (a) Optical microscope image of the mechanically exfoliated monolayer graphene flake with gold electrodes. The wiring of the STM tip and back gate voltage is indicated.  (b) STM topographic image of monolayer graphene on hBN showing the underlying surface corrugations. The image is 100 nm x 100 nm. The imaging parameters are tip voltage $V_t$ = -0.3 V, tunneling current $I_t$ = 100 pA. (c) STM topographic image of monolayer graphene on SiO$_2$ showing significantly increased corrugations. The imaging parameters are tip voltage $V_t$ = -0.5 V, tunneling current $I_t$ = 50 pA. (d) Histogram of the height distributions for graphene on SiO$_2$ (blue squares) and graphene on hBN (red triangles) along with gaussian fits. \label{topography}}
\end{figure}

To study how the local electronic structure of graphene is affected by the hBN substrate, we prepare graphene on hBN devices for STM measurements.  A schematic of the measurement set-up showing the graphene flake on hBN with gold electrodes for electrical contact is shown in Fig. \ref{topography}a.  A typical STM image of the monolayer graphene showing the surface corrugations due to the underlying hBN substrate is shown in Fig. \ref{topography}b.  This image can be compared with an STM image of monolayer graphene prepared in a similar manner but on SiO$_2$ as shown in Fig. \ref{topography}c.  It is clear from these two images that the surface corrugations are much larger for graphene on SiO$_2$ as compared to hBN.  This is due to the graphene conforming to the substrate and the planar nature of hBN as compared to the amorphous SiO$_2$.  Figure \ref{topography}d shows a histogram of the heights in the two images.  In both cases, the heights are well described by gaussian distributions with standard deviations of 224.5 $\pm$ 0.9 pm for graphene on SiO$_2$ and 30.2 $\pm$ 0.2 pm for graphene on hBN.  The values for graphene on SiO$_2$ are similar to previously reported values \cite{Ishigami2007, Stolyarova2007} while the distribution for graphene on hBN is similar to graphene on mica or HOPG\cite{Lui2009}.  Reducing the surface roughness is critical for graphene devices because local curvature can lead to electronic effects such as doping\cite{Kim2009} and random effective magnetic fields\cite{Guinea2008}.  As the height variation of the graphene on hBN is as flat as HOPG, it has reached its ultimate limit of flatness.

\begin{figure}[h]
\includegraphics[]{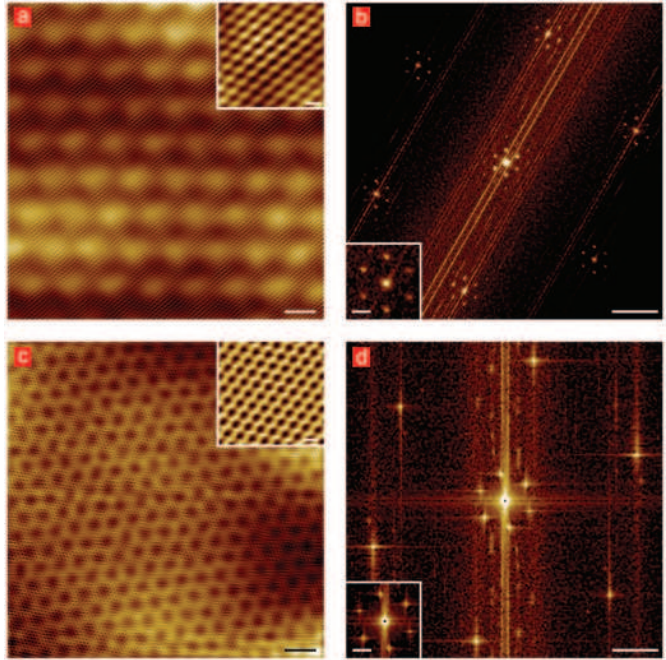}
\caption{Real space and Fourier transforms of Moir\'{e} patterns (a) STM topography images of a Moir\'{e} pattern produced by graphene on hBN. The scale bar is 2 nm.  The inset is a zoom in of a 2 nm region and the scale bar is 0.3 nm.  The imaging parameters are V$_t$ = -0.3 V and I$_t$ = 100 pA.  (b) Fourier transform of (a) showing the six graphene lattice points near the edge of the image and the long wavelength Moir\'{e} pattern near the center of the image and around each lattice point.  The scale bar is 10 nm$^{-1}$.  The inset is a zoom in around one of the lattice points.  Its scale bar is 2 nm$^{-1}$. (c) STM topography image from another region of the same graphene flake showing a different Moir\'{e} pattern.  The scale bar is 2 nm.  The inset is a zoom in of a 2 nm region and the scale bar is 0.3 nm.  The imaging parameters are V$_t$ = -0.3 V and I$_t$ = 100 pA.  (d) Fourier transform of (c) showing the atomic lattice as well as the Moir\'{e} pattern.  The scale bar is 10 nm$^{-1}$.  The inset is a zoom of the Moir\'{e} pattern and its scale bar is 4 nm$^{-1}$. \label{FFT}}
\end{figure}

Looking more closely at the topography of the graphene on hBN, we resolve its atomic lattice and also observe longer periodic modulations which result in a distinct Moir\'{e} pattern as seen in Figs. \ref{FFT}a and \ref{FFT}c.  These two images were acquired from different areas of the same graphene flake and show Moir\'{e} patterns of different length due to a shift in the alignment of the graphene with the underlying hBN lattice.  In the case of Figs. \ref{FFT}a and \ref{FFT}c, we find a long wavelength modulation of 2.6 nm and 1.3 nm respectively.  By examining the Fourier transform of these images, we can learn about the underlying hBN substrate as well as its relative orientation with respect to the graphene lattice.  In both cases, we see two distinct sets of peaks in the FFT, there is a set of six points near the center of the images which correspond to the Moir\'{e} pattern and there are six additional points near the edge of the images which correspond to the graphene atomic lattice.  The first observation is that the locations of these points is rotated with respect to each other in the two images.  For Fig. \ref{FFT}b, the atomic lattice is rotated by 12.6 $\pm$ 1.0$^\circ$ from the horizontal.  In the case of Fig. \ref{FFT}d, the atomic lattice is rotated by 18.5 $\pm$ 0.6$^\circ$.  In contrast the Moir\'{e} pattern is rotated by 29.9 $\pm$ 0.1$^\circ$ from the horizontal in Fig. \ref{FFT}b and 38.4 $\pm$ 0.2$^\circ$ from the horizontal in Fig. \ref{FFT}d.  From the lengths of the Moir\'{e} patterns and their angles, we can calculate the orientation of the graphene lattice with respect to the underlying hBN (see Supplementary Information).  In the case of Fig. \ref{FFT}a, we find that the hBN is rotated by -5.4$^\circ$ from the graphene.  On the other hand, for Fig. \ref{FFT}c, we find that the hBN is rotated by -10.9$^\circ$.  This difference of 5.5$^\circ$ matches the difference in the orientation of the graphene lattices in the two images.  Therefore, we conclude that the underlying hBN substrate is continuous and the graphene above it sits at two different angles.  Atomic force microscopy images show that graphene on hBN tends to form flat regions separated by ridges and pyramids.  As the two STM images were taken from different sides of one of these ridges, it is clear that the graphene can change orientation across these ridges.

\begin{figure}[h]
\includegraphics[]{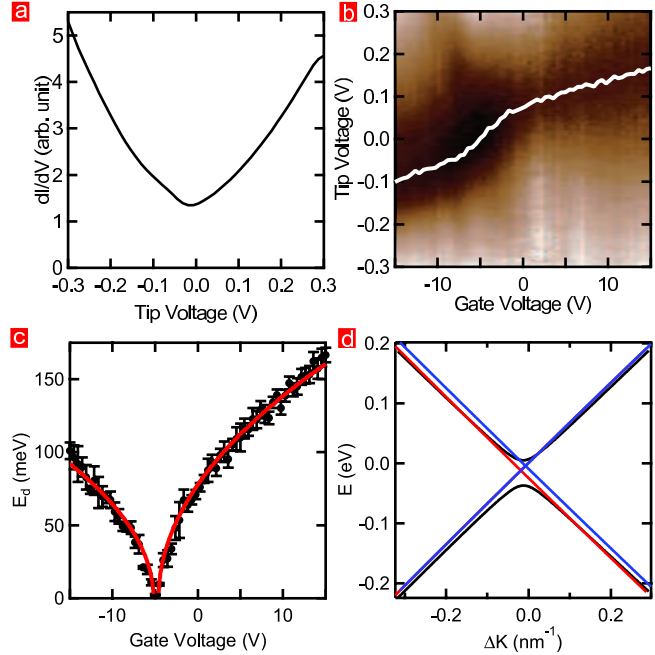}
\caption{Spectroscopy of graphene on hBN as a function of gate voltage (a) dI/dV spectroscopy showing a nearly linear density of states as a function of energy (tip voltage).  (b) dI/dV spectroscopy as a function of tip voltage and gate voltage.  The white line corresponds to the minimum in the dI/dV curves and represents the Dirac point. (c) Energy of the Dirac point as a function of gate voltage.  The red curve is a fit assuming a linear band structure. (d) Energy versus momentum dispersion relations for the case of graphene and hBN having the same lattice constant and zero angle mismatch (black curve) and two curves with 1.8\% lattice mismatch.  The blue curve has -5.45$^{\circ}$ angle mismatch and the red curve has -10.9$^{\circ}$.  \label{FermiVelocity}}
\end{figure}

The scanning tunneling microscope is not only able to acquire images of atoms but can also map the local density of states.  We have performed scanning tunneling spectroscopy of the graphene on hBN.  Figure \ref{FermiVelocity}a shows a typical dI/dV spectroscopy curve which is proportional to the local density of states.  The curve is nearly linear in energy with a minimum near zero tip voltage indicating the energy of the Dirac point.  The location of this minimum can be varied by applying a back gate voltage, $V_g$, to the sample.  The voltage on the back gate induces a charge on the graphene of $n = \alpha V_g$ with $\alpha = 7.2 \times 10^{10}$ e/cm$^2$V based on a parallel plate capacitor model.  In the model, we have taken 285 nm of SiO$_2$ with a dielectric constant of 3.9 and 14 nm of hBN with a dielectric constant of 3-4\cite{Young2010}. There is a small uncertainty in the value of $\alpha$ of about 1\% due to the unknown precise value of the hBN dielectric constant.  Figure \ref{FermiVelocity}b plots dI/dV as a function of tip voltage and gate voltage.  The white line follows the minimum in the dI/dV curves for each value of the gate voltage.  This minimum occurs when the Fermi energy of the tip lines up with the Dirac point.  We observe that the location of the minimum changes more quickly when the Dirac point is near zero tip voltage which is consistent with the linear band structure of graphene.  There is also a dark ridge that occurs at decreasing tip voltage as the gate voltage is increased.  This is due to the effect of the voltage on the tip acting as a local gate and changing the density of electrons in the graphene\cite{Choudhury2010}.

Our local spectroscopy measurements indicate that there is no band gap induced in graphene on hBN, not even locally. These results disagree with earlier theoretical calculations which predicted the opening of a band gap of order 50 meV when graphene is placed on hBN, because of the breaking of sublattice symmetry\cite{Giovannetti2007, Slawinska2010}. This discrepancy is explained by the 1.8\% mismatch between the graphene and hBN lattices and the different orientations of the two lattices, which were both neglected in Refs.~\cite{Giovannetti2007, Slawinska2010}. Taking them into account, one expects that, while one of the carbon atoms may sit over a boron (nitrogen) atom at one location, this alignment gets lost a few lattice constants away. In large enough systems, carbon atoms should therefore have the same probability to have a boron or a nitrogen atom as nearest neighbor in the hBN layer, regardless of their sublattice index. We numerically checked the validity of this hypothesis by calculating the interlayer hopping potential  $\sim \gamma_\perp \exp[-|{\bf r}-{\bf r}'|/\xi]$ from a carbon atom at ${\bf r}$ in the graphene layer to a boron or nitrogen atom at ${\bf r}'$ in a rotated hBN layer. We restricted ourselves to nearest- and next-nearest-neighbor interlayer hopping and chose the parameters $\gamma_\perp = 0.39$ eV and $\xi=0.032$ nm to fit known values for these hoppings in graphene bilayers\cite{Zhang2008}. We found that the 1.8\% lattice mismatch alone is sufficient to make the hopping strength from a carbon atom to a boron or a nitrogen atom independent of the graphene sublattice index for systems of a few hundred unit cells, going down to a few tens of unit cells when the lattices are misaligned by about one degree. We incorporated the Fourier transform of this hopping potential into the low-energy Hamiltonian for graphene on hBN to find the energy-momentum dispersion. The inter-layer coupling is nonzero only for ${\bf k}=0$ as well as for six additional vectors ${\bf k}$ associated with the Moir\'{e} pattern. Most importantly, we found that the coupling between the A and B atoms in the graphene lattice with the boron and nitrogen atoms in the hBN are almost identical. Thus sublattice symmetry is restored and a gapless Dirac spectrum is recovered, albeit at slightly shifted values of ${\bf K}$. This is illustrated in Fig.~\ref{FermiVelocity}d. While $\gamma_\perp$ depends in principle on whether hopping occurs between a carbon and a boron or nitrogen atom, we note that this does not break sublattice symmetry, and we checked that the spectrum remains gapless, even when this discrepancy is taken into account. More details of our numerical approach are given in the Supplementary Information. 
  
By determining the energy of the Dirac point as a function of gate voltage, we can measure the Fermi velocity of electrons and holes in graphene.  Figure \ref{FermiVelocity}c shows the energy of the Dirac point as a function of gate voltage.  Graphene has a linear dispersion relation such that $E = \hbar v_F k$ where $v_F$ is the Fermi velocity.  Since it is a two-dimensional material, the density of electrons is given by $n = g_s g_v \pi k^2/(2\pi)^2$ where $g_s$ and $g_v$ are the spin and valley degeneracy which are both 2 for graphene.  Therefore, the Dirac point should depend on gate voltage as $E = \hbar v_F \sqrt{\pi \alpha V_g}$.  The red curve is a fit to the data from which we can extract the Fermi velocity.  We find that $v_F = 1.16 \pm 0.01 \times 10^6$ m/s for the electrons and $v_F = 0.94 \pm 0.02 \times 10^6$ m/s for the holes.  Moreover, we observe an asymmetry between the Fermi velocity for electrons and holes of about 25\% depending on the Moir\'{e} pattern observed.  The shorter Moir\'{e} pattern has a higher Fermi velocity for holes while the longer one has a higher Fermi velocity for electrons.  The origin of this asymmetry is unclear but may arise due to next-nearest neighbor coupling which are not taken into account in our model. 

\begin{figure}[h]
\includegraphics[]{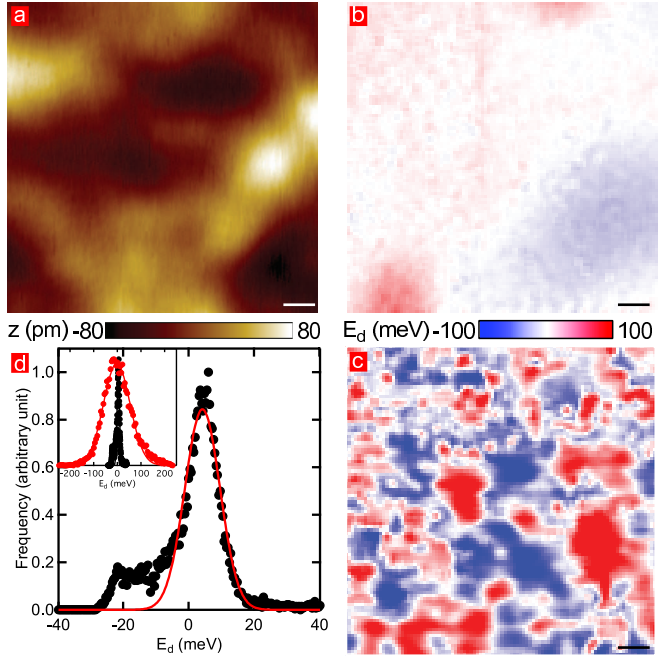}
\caption{Spatial maps of the density of states of graphene on hBN and SiO$_2$ (a) Topography of graphene on hBN. (b) Tip voltage at the Dirac point as a function of position for graphene on hBN. (c) Tip voltage at the Dirac point as a function of position for graphene on SiO$_2$.  The color scale is the same for (b) and (c). (d) Histogram of the energies of the Dirac point from (b) as well as a gaussian fit.  The inset shows the same data but also includes the histogram for SiO$_2$ shown in red.  The scale bar in all images is 10 nm. \label{SpatialMap}}
\end{figure}

One of the main advantages to using hBN as a substrate for graphene as compared to SiO$_2$ is the improvement in electronic properties of the graphene which is believed to be due to the lack of charge traps on the hBN surface.  Figure \ref{SpatialMap}a shows the topography of graphene on hBN over a range of 100 nm.  Note that the height variation is less than 0.1 nm over the range of the image as compared to typical values of nearly 1 nm for graphene on SiO$_2$.  We have performed dI/dV measurements at 1 nm intervals over the entire area of Fig. \ref{SpatialMap}a.  For each of these dI/dV curves, we have found tip voltage of the minimum which corresponds to the Dirac point.  The results are plotted in Fig. \ref{SpatialMap}b.  We have done a similar analysis for a 100 nm area of graphene on SiO$_2$ and the results are plotted in Fig. \ref{SpatialMap}c.  The red and blue regions correspond to electron and hole puddles respectively.  It is clear from these two images that the variation in the energy of the Dirac point is much smaller on hBN.  The spatial extent of each puddle is also much smaller in the graphene on SiO$_2$ consistent with an increased density of impurities\cite{Rossi2008}.

We can further quantify the disorder in the graphene by looking at a histogram of the energy of the Dirac point, Fig. \ref{SpatialMap}d.  The main part of the histogram for the Dirac point energy on hBN is well fit by a gaussian distribution (red line) with a standard deviation of $5.4 \pm 0.1$ meV.  In addition, there is a small extra bump in the distribution from the hole doped region near the bottom right of Fig. \ref{SpatialMap}b.  In comparison, the distribution on SiO$_2$ is much broader with a standard deviation of $55.6 \pm 0.7$ meV.  These distributions in energy can be converted to charge fluctuations using $n = E_d^2/\pi(\hbar v_F)^2$.  We find that the charge fluctuations in graphene on hBN are $\sigma_n = 2.50 \pm 0.13 \times 10^9$ cm$^{-2}$ while they are more than 100 times larger for graphene on SiO$_2$, $\sigma_n = 2.64 \pm 0.07 \times 10^{11}$ cm$^{-2}$.  Our measurements for the charge fluctuations on SiO$_2$ are consistent with previous single electron transistor\cite{Martin2008} and STM\cite{Deshpande2009, Zhang2009} measurements which established the presence of electron and hole puddles in graphene on SiO$_2$.  Furthermore, our measurements for the charge fluctuations in graphene on hBN show a very similar value to values extracted from electrical transport measurements in suspended graphene samples\cite{Du2008} implying that using hBN as a substrate provides a similar benefit to suspending graphene without the associated fabrication challenges and limitations.

We have demonstrated that graphene on hBN provides an extremely flat surface that has significantly reduced electron-hole puddles as compared to SiO$_2$.  By reducing the charge fluctuations, the low density regime and the Dirac point can be more readily accessed.  Moreover, hBN allows this low-density regime to be reached in a substrate-supported system which will allow atomic resolution local probes studies of the Dirac point physics.

\section*{Methods}
Thin and flat few layer hBN flakes were prepared by mechanical exfoliation of hBN single crystals on SiO$_2$/Si substrates.  The hBN growth method has been previously described\cite{Taniguchi2007}.  Exfoliated graphene flakes were then transferred to the hBN using Poly(methyl methacrylate) (PMMA) as a carrier \cite{Dean2010}. Then Cr/Au electrodes were deposited using standard electron beam lithography. The lithography process leaves some PMMA resist on the surface of graphene which is cleaned by annealing in argon and hydrogen at 350$^\circ$ C for 3 hours \cite{meyer2007}. The device was then immediately transferred to the STM (Omicron low temperature STM operating at T = 4.5 K in ultrahigh vacuum (p $\leq 10^{-11}$ mbar)).  Electrochemically etched tungsten tips were used for imaging and spectroscopy.  All of the tips used were first checked on an Au surface to ensure that their density of states was constant.

The dI/dV spectroscopy was acquired by turning off the feedback loop and holding the tip a fixed distance above the surface.  A small ac modulation of 5 mV at 563 Hz was applied to the tip voltage and the corresponding change in current was measured using lock-in detection.  We also measured dI/dV curves with 0.5 mV excitation and observed the same results.


\section*{acknowledgments}
The authors would like to thank W. Bao, Z. Zhao and C.N. Lau for providing the graphene on SiO$_2$ samples used for the comparison with graphene on hBN.  J.X., A. D. and B.J.L. were supported by U. S. Army Research Laboratory and the U. S. Army Research Office under contract/grant number W911NF-09-1-0333 and the National Science Foundation CAREER award DMR-0953784.  P.J. was supported by the National Science Foundation under award DMR-0706319.  J.S.-Y., D.B. and P. J.-H. were supported by the U.S. Department of Energy, Office of Basic Energy Sciences, Division of Materials Sciences and Engineering under Award \#DE-SC0001819 and by the 2009 U.S. Office of Naval Research Multi University Research Initiative (MURI) on Graphene Advanced Terahertz Engineering (Gate) at MIT, Harvard and Boston Unversity.

\section*{Author contributions}
J.X. and B.J.L. performed the STM experiments of the graphene on hBN.  A.D. performed the STM experiments of graphene on SiO$_2$.  J. S.-Y. and D. B. fabricated the devices.  P.J. performed the theoretical calculations.  K.W. and T.T. provided the single crystal hBN.  P. J.-H. and B.J.L. conceived and provided advice on the experiments. All authors participated in the data discussion and writing of the manuscript.

\section*{Competing financial interests}
The authors declare no competing financial interests.

\setcounter {figure} {0}
\renewcommand{\thefigure}{S\arabic{figure}}
\section*{Supplementary Information}
\section{Moir\'{e} Patterns}
A Moir\'{e} pattern occurs when the atoms in the graphene layer form a super-lattice structure with the atoms in the hBN layer.  In this section, we derive the conditions which lead to a Moir\'{e} pattern and therefore predict the angles and lengths of the Moir\'{e} pattern.

We consider two superimposed hexagonal lattices defined by 
\begin{equation}
{\bf r}_{i \mu}(m,n) = n {\bf a}_{i+} + m {\bf a}_{i-} +(\mu - i) {\bf d},
\end{equation}
with the layer index $i=1,2$ for the graphene and hBN lattices, respectively, 
and the sublattice index $\mu=1$ (A sublattice), 2 (B sublattice). 
In the hBN layer, $\mu=1$ for nitrogen and 2 for boron. We define
${\bf d}=(0,-1/\sqrt{3}) a_i$ as the vector connecting the two sublattices, with lattice spacing $a_i$. The vectors lattice vectors associated with the hBN, ${\bf a}_{2\pm}$, are rotated by an angle $\phi$ counterclockwise with respect to the graphene lattice ${\bf a}_{1\pm} = (\pm 1/2,\sqrt{3}/2) a_1$ and are longer by a factor  $a_2/a_1= 1.018$.  Therefore the hBN lattice vectors are given by ${\bf a}_{2\pm} = (\pm\cos(\pi/3 \pm \phi), \sin(\pi/3 \mp \phi)) a_2$

An arbitrary point in the graphene lattice can be represented by two integers $(n,m)$ which correspond to the number of lattice vectors in the ${\bf a}_{1+}$ and ${\bf a}_{1-}$ directions respectively.  Similarly, a position in the hBN lattice is represented by the integers $(r,s)$.  If the graphene and hBN lattice are AB stacked at a given position, they will be AB stacked again when $n {\bf a}_{1+} + m {\bf a}_{1-} = r {\bf a}_{2+} + s {\bf a}_{2-}$.  In terms of $xy$-coordinates this gives the conditions
$$n - m = 2\frac{a_2}{a_1} (r \cos(\pi/3+\phi) - s \cos(\pi/3-\phi))$$
$$\sqrt{3}(n + m) = 2\frac{a_2}{a_1} (r\sin(\pi/3-\phi) + s \sin(\pi/3+\phi))$$
A Moir\'{e} pattern occurs when these equations can be satisfied for integer values of $(n,m)$ and $(r,s)$.  For a given ratio of $a_2/a_1$ these conditions will only hold for certain special angles.  In terms of the values of $(n,m)$, the length of the Moir\'{e} pattern can be written as $L = a_1\sqrt{n^2+m^2+nm}$ and it is at an angle $\theta = \tan^{-1}(\frac{\sqrt{3}m}{2n+m})$ with respect to ${\bf a}_{1+}$.  However, a near commensurate condition can always be found leading to a Moir\'{e} pattern over some finite length.

\begin{figure}[h]
\includegraphics[]{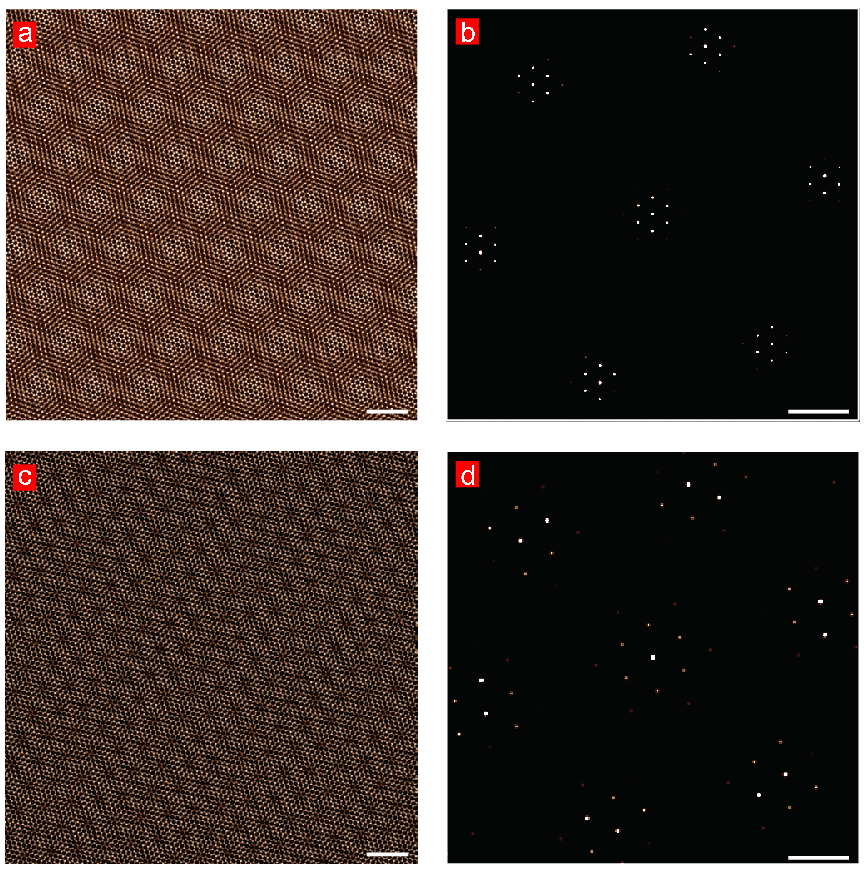}
\caption{Simulated real space and Fourier transforms of Moir\'{e} patterns (a) Simulated lattice showing a Moir\'{e} pattern produced by graphene on hBN. The scale bar is 2 nm.  (b) Fourier transform of (a) showing the six graphene lattice points near the edge of the image and the long wavelength Moir\'e pattern near the center of the image and around each lattice point.  The scale bar is 10 nm$^{-1}$.  (c) Simulated lattice showing a shorter Moir\'{e} pattern.  The scale bar is 2 nm.  (d) Fourier transform of (c) showing the atomic lattice as well as the Moir\'{e} pattern. The scale bar is 10 nm$^{-1}$}  \label{SIfig}
\end{figure}

We have numerically created lattices to reproduce the images in Figure 2.  The results are shown in Fig. \ref{SIfig}.  The first set of images correspond to the data shown in Fig. 2a and 2b in the main text.  It was created by using a hBN lattice with a 1.8\% longer lattice constant and rotated by $\phi=-5.4^{\circ}$ counterclockwise with respect to the graphene lattice.  From the FFT of the lattice, Fig. \ref{SIfig}b, we see that it matches the experimental figure, Fig. 2b very well.  To create the shorter Moir\'{e} pattern we must use a larger rotation angle.  We find the best match occurs at $\phi=-10.9^{\circ}$.  In general, the Moir\'{e} pattern gets shorter as the angle $\phi$ increases.

\section{Theory Calculations}

We restrict the interlayer hopping potential to nearest-neighbor and next-nearest-neighbor hopping, and evaluate
\begin{eqnarray}
V_{\mu \nu} (m,n) & = & \gamma_\perp \exp[-|{\bf r}_{1 \mu}(m,n)-{\bf r}_{2 \nu}(m',n')|/\xi] \, ,
\end{eqnarray}
with $(m',n')$ labelling the nearest and next-nearest neighbor sites to $(m,n)$. Note that if one is a boron atom, the other one is a nitrogen atom. In this way, four 
interlayer couplings between the two sublattices are defined. The parameters $\gamma_\perp=0.39$ eV and $\xi=0.032$ nm are calibrated to fit the interlayer couplings in bilayer graphene\cite{Zhang2008}. While $\gamma_\perp$ should in 
principle depend on the sublattice index $\mu$ in the hBN layer, this has no influence on our main conclusion,
that the graphene spectrum is effectively gapless due to the mismatch between lattices and their different relative
orientations, because it does not break sublattice symmetry in the graphene layer. 

\begin{figure}[h]
\includegraphics[]{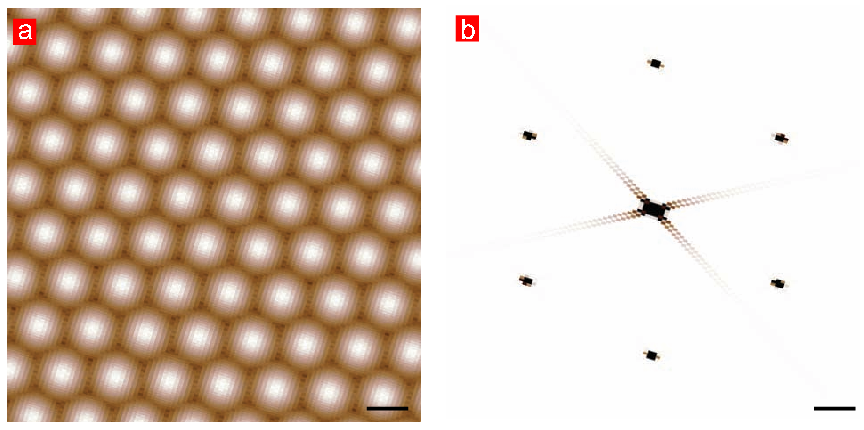}
\caption{Real space and Fourier transforms of the interlayer hopping (a) $V_{AA}$ potential for an interlayer rotation of $\phi=-5.4^o$. The scale bar is 2 nm.  (b) Fourier transform of (a) showing a central peak at ${\bf k}_0=0$ and six additional peaks due to the Moir\'e pattern.  The scale bar is 0.8 nm$^{-1}$.  }  \label{Vmunu}
\end{figure}

We consider system sizes up to $600 \times 600$ unit cells, 
which is more than sufficient to extract the periodicity of the Moir\'e patterns shown in Figs. 2a and 2c.
Fig.~\ref{Vmunu}a
shows a color plot of $V_{AA} (m,n)$ for an interlayer rotation angle of $\phi=-5.4^o$ corresponding to 
the sample shown in Fig. 2a. The potential has a clear hexagonal periodicity, with a wavelength of about 10 lattice sites, in quantitative agreement with Fig. 2a. From that hexagonal pattern, the Fourier transform $\tilde{V}_{\mu \nu}({\bf k})$ 
of  $V_{\mu \nu} (m,n)$ generically exhibits seven peaks, including a 
large one at ${\bf k}_0=0$, and six secondary peaks 
at wave vectors ${\bf k}_i$, $i=1,\ldots 6$ corresponding to the 
Moir\'e pattern. 
This is shown in Fig.~\ref{Vmunu}b, with hexagonally
distributed secondary peaks at positions reflecting the central pattern of Fig. 2b.

The hopping potential induces interlayer coupling which we introduce in the low-energy Hamiltonian. We keep only the
seven dominant transitions and because the hBN bands are several eV's away from the Dirac points and $\tilde{V}_{\mu \nu} ({\bf k}_{i \ne 0})/\tilde{V}_{\mu \nu} ({\bf k}_0) \lesssim 0.1$, we truncate the Hamiltonian to a 16 $\times$ 16 matrix
\begin{eqnarray}
{\cal H}({\bf q},{\bf q}') &=& \left[ H_{\rm C}({\bf q}) + \sum_{i=0}^6 H_{\rm BN}({\bf q}+{\bf k}_i)\right]
\delta_{{\bf q},{\bf q}'} + \left[\sum_{i=0}^6 H_{\rm C,BN}({\bf k}_i) + h.c. \right] \delta_{{\bf q}+{\bf k}_i,{\bf q}'} \, ,
\end{eqnarray}
with all $H$'s being $2 \times 2$ matrices, 
\begin{subequations}
\begin{eqnarray}
H_{\rm C}({\bf q}) &=& \left(
\begin{array}{cc}
0 & \gamma_0 f_0({\bf q}) \\
\gamma_0 f^*_0({\bf q}) & 0 
\end{array}
\right) \, ,\\
H_{\rm BN}({\bf q}) &=& \left(
\begin{array}{cc}
\epsilon_{\rm B} & \gamma_1 f_1({\bf R^{-1}(\phi) \, q}) \\
\gamma_1 f^*_1({\bf R^{-1}(\phi) \, q}) & \epsilon_{\rm N}
\end{array}
\right) \, ,\\
H_{\rm C,BN}({\bf k}_i)&=& \left(
\begin{array}{cc}
\tilde{V}_{AA}({\bf k}_i) &\tilde{V}_{AB}({\bf k}_i) \\
\tilde{V}_{BA}({\bf k}_i) & \tilde{V}_{BB}({\bf k}_i)
\end{array}
\right) \, ,
\end{eqnarray}
\end{subequations}
with $f_i({\bf q})=1 + 2 \cos(q_x a_i/2) \exp(-i \sqrt{3} q_y a_i/2)$
and parameters $\gamma_0=3.16$ eV, $\gamma_1=2.79$ eV, $\epsilon_{\rm B}=3.34$ eV and
$\epsilon_{\rm N}=-1.4$ eV\cite{Slawinska2010}. The low-energy graphene dispersion is obtained via diagonalization
of ${\cal H}$. We find that a gap exists in the spectrum only 
when $\tilde{V}_{\mu {\rm A}}({\bf k}_i)-\tilde{V}_{\mu {\rm B}}({\bf k}_i) \ne 0$ for at least one $\mu$, thereby breaking sublattice symmetry
in the graphene layer. 
At experimentally relevant rotation angles $\phi$ of a few degrees, we find that ${\rm max} [\tilde{V}_{\mu \nu} ({\bf k}_i)-\tilde{V}_{\mu' \nu'} ({\bf k}_i)]/ \tilde{V}_{\mu \nu} ({\bf k}_i) \lesssim 10^{-4} $ for 150 $\times$ 150 systems, decreasing with size to less than
$10^{-5}$ for 600 $\times$ 600 systems. Consequently, we get an upper bound of $\Delta < 10^{-6}$ eV for the
excitation gap $\Delta$ between the two graphene bands for the largest systems
we investigated. The latter are still smaller than the experimentally 
observed domains so that such a small gap cannot be resolved.

Large values for ${\rm max} [\tilde{V}_{\mu \nu} ({\bf k}_i)-\tilde{V}_{\mu' \nu'} ({\bf k}_i)]/ \tilde{V}_{\mu \nu} ({\bf k}_i)$ are obtained only for
small systems or if the lattice mismatch is neglected. This explains the
50 meV gap reported in Refs.~\cite{Giovannetti2007, Slawinska2010}, which we
qualitatively reproduced (see Fig. 3d). Even without relative rotation of the two layers, we found that the lattice mismatch is
sufficient to effectively close this gap already for systems of sizes 300 $\times$ 300. We finally note that the expected different hoppings on the boron and nitrogen atoms induce differences in $\tilde{V}_{AA}$ vs. $\tilde{V}_{AB}$ as well as in $\tilde{V}_{BA}$ vs. $\tilde{V}_{BB}$. However, such differences do not break sublattice symmetry and therefore do not open a gap. 

\end{document}